\documentclass[ 
aip, jcp,
groupedaddress,
amssymb,
amsmath,
 floatfix,
 reprint,
 tightenlines,
 longbibliography]{revtex4-2}

\usepackage{pgfplots} 
\pgfplotsset{compat=1.18}
\usepackage{graphicx}
\usepackage{dcolumn}
\usepackage{bm}
\usepackage{bbm}
\usepackage{siunitx}
\definecolor{linkcolor}{rgb}{0.3,0.3,1.0} 
\usepackage[pdftex,colorlinks=true, linkcolor= linkcolor, citecolor= linkcolor, urlcolor= linkcolor, hyperindex=true,hyperfigures=true]{hyperref}

\usepackage[english]{babel}

\newcommand*{\e}{\mathrm{e}}

\renewcommand*{\i}{\mathrm{i}}
\newcommand*{\abs}[1]{\left\lvert{#1}\right\rvert}
\providecommand*{\avg}[1]{\left\langle#1\right\rangle}

\renewcommand*{\vec}[1]{\bm{#1}}

\makeatletter
\renewcommand*{\d}%
{\@ifnextchar^{\DIfF}{\DIfF^{}}}
\def\DIfF^#1{%
	\mathop{\mathrm{\mathstrut d}}%
	\nolimits^{#1}\gobblespace
}
\def\gobblespace{%
	\futurelet\diffarg\opspace}
\def\opspace{%
	\let\DiffSpace\!%
	\ifx\diffarg(%
	\let\DiffSpace\relax
	\else
	\ifx\diffarg\[%
	\let\DiffSpace\relax
	\else
	\ifx\diffarg\{%
	\let\DiffSpace\relax
	\fi\fi\fi\DiffSpace}

\DeclareMathOperator{\VV}{\mathcal{U}}

\newcommand{\UNIPD}{
Department of Physics and Astronomy, University of Padova, Via Marzolo 8, I-35131 Padova, Italy}
\newcommand{\INFN}{INFN, Sezione di Padova, Via Marzolo 8, I-35131 Padova, Italy}

\begin{document}
\title{Fluctuations of driven probes reveal nonequilibrium transitions in complex fluids}
\author{Danilo Forastiere}
\author{Emanuele Locatelli}
\author{Gianmaria Falasco}
\author{Enzo Orlandini}
\author{Marco Baiesi}
\email{marco.baiesi@unipd.it}
\affiliation{\UNIPD}
\affiliation{\INFN}

\date{\today}

\begin{abstract}
Complex fluids subjected to localized microscopic energy inputs, typical of active microrheology setups, exhibit poorly understood nonequilibrium behaviors because of the intricate self-organization of their mesoscopic constituents.
In this work we show how to identify changes in the microstructural conformation of the fluid by monitoring the variance of the probe position, based on a general method grounded in the breakdown of the equipartition theorem.
To illustrate our method, we perform large-scale Brownian dynamics simulations of an effective model of micellar solution, and we link the different scaling regimes in the variance of the probe's position to the transitions from diffusive to jump dynamics, where the fluid intermittently relaxes the accumulated stress. 
This suggests stored elastic stress may be the physical mechanism behind the nonlinear friction curves recently measured in micellar solutions, pointing at a mechanism for the observed multi-step rheology. 
Our approach overcomes the limitations of continuum macroscopic descriptions and introduces an empirical method, applicable in experiments, to detect nonequilibrium transitions in the structure of complex fluids.
\end{abstract}

\maketitle

\section{Introduction}

Considerable effort has been devoted to understanding complex fluids, including polymers and micellar networks~\cite{gelbart1996new, cates1990statics, cates2017complex}, as well as active soft fluids~\cite{marchetti2013hydrodynamics, cates2015motility, fodor2016far}. 
Recent experimental studies using optical tweezers~\cite{scott2023extracting, volpe2023roadmap} have revealed surprising phenomena such as oscillations and recoil of overdamped particles~\cite{berner2018oscillating, ginot2022recoil}. 
Furthermore, the connection between local microstructure and mechanical behavior has been extended to active matter~\cite{paul2022force, knippenberg2024motility}. 
Most of these studies focused on the average nonlinear mechanical response.
 In analogy to the equilibrium case,  where fluctuations and response are linked by the fluctuation-dissipation theorem~\cite{marconi2008fluctuation}, the question arises of characterizing the fluctuations typical of these setups, which necessarily bear the hallmark of nonequilibrium~\cite{muenker2024accessing, di2024variance}.

The components of complex fluids such as polymers or micelles self-organize in highly hierarchical ways in thermodynamic equilibrium~\cite{doi1988theory, cates1990statics, cates2017complex}, creating mesoscopic structures spanning multiple length scales, sometimes differing by orders of magnitude, which may result in prolonged relaxation of some degrees of freedom~\cite{iubini2020aging,baiesi2021rise}.
Therefore, localized perturbations made through optical tweezers may bring the fluid locally far from thermodynamic equilibrium. 
As a consequence, the universal link between fluctuations and response, and the energy equipartition among degrees of freedom, break down~\cite{baiesi2013update, falasco2016mesoscopic}, and more detailed modeling of the experimental situation is called for. 

The setting on which we focus is that of microrheology~\cite{mason1997particle, waigh2005microrheology, waigh2016advances,puertas2014microrheology, furst2017microrheology, zia2018active, robertson2018optical}, used to measure the properties of complex fluids by tracking and interpreting the motion of mesoscopic probes. 
\emph{Passive} microrheology typically uses information on a free probe's trajectories to determine the small-amplitude, frequency-dependent response moduli of equilibrium fluids (via the fluctuation-dissipation theorem and the Stokes-Einstein relation)~\cite{levine2000one, waigh2016advances, furst2017microrheology} or nonequilibrium ones~\cite{muenker2024accessing}.
Conversely, {\em active} microrheology relies on the average response of a particle to an external force. It conceptually resembles macroscopic rheology, which measures the stress from imposing a predetermined flow pattern on a sample and provides access to the nonlinear features of the response~\cite{ squires2005simple, squires2008nonlinear, gazuz2009active, furst2017microrheology, zia2018active, robertson2018optical, ginot2022recoil, caspers2023mobility, paul2022force}.
Applying a known external force to a probe particle via optical tweezers, one obtains a direct measure of the friction coefficient seen by the probe, which near equilibrium can be related to the bulk viscosity of the solution~\cite{batchelor2000introduction, squires2005simple}. 
However, the connection between the probe's observed motion 
(\emph{i.e.}, the friction curves as a function of the velocity) 
and the underlying microscopic dynamics of the fluid remains elusive in the case of complex fluids in the nonlinear regime~\cite{berner2018oscillating, ginot2022recoil, caspers2023mobility}, even in the simple setting of a probe dragged at a constant speed in a micellar fluid or polymer solution~\cite{chapman2014nonlinear, falzone2015entangled, jain2021two, peddireddy2022optical}.
Specifically, recent work~\cite{jain2021two} showed that the localized driving protocol leads to multistep curves for nonlinear friction and to discrepancies between microscopic and macroscopic viscosity, as also observed in colloidal suspensions~\cite{squires2005simple, furst2017microrheology},
pointing to the onset of noncontinuum effects~\cite{chapman2014onset, falzone2015entangled}, where the size and conformation of the fluid's mesoscopic constituents are relevant.

We aim to link the fluctuations in the position of the probe with the hidden microscopic state of the fluid, and to use this connection to shed light on the mechanical properties of the fluid.
This work introduces an approach to infer changes in the nonequilibrium microstructure of complex fluids that display a diverse array of nonlinear behaviors when subjected to the motion of a probe dragged by traveling optical tweezers.
We achieve this by identifying transitions between different scaling regimes in the probe's fluctuations, a practical and already accessible quantity, yet crucially not usually exploited in current microrheological setups.
Our proposed method stems from a theoretical analysis and detailed numerical simulations of a mesoscopic model for polymeric fluids.
The scaling of the steady-state variance of the probe's position as a function of the drag speed (when properly adimensionalized) allows us to distinguish between near-equilibrium and different nonequilibrium regimes. In our numerical model, we link these transitions to the spatial patterns of polymer deformation and to the onset of an activated jump dynamics for stress relaxation, beyond a threshold velocity $v^*$.

\begin{figure}
\centering
\includegraphics[scale=0.4]{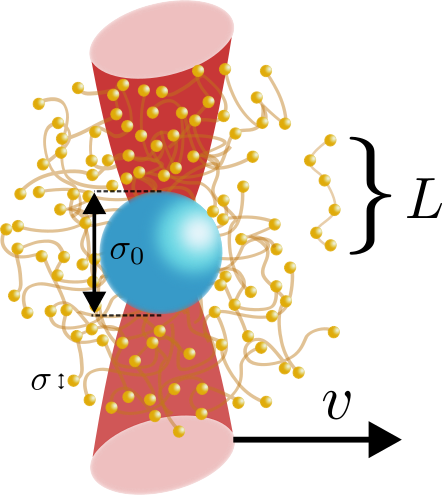}
 \caption{ 
 Schematic representation of our two-dimensional simulation setup. 
 A probe of nominal diameter $\sigma_0$ interacts via soft Gaussian repulsion with  harmonic chains, with fixed length $L$, of units of diameter $\sigma=\sigma_0/10$, which repel each other via a soft Gaussian potential. 
 The probe is driven by a harmonic potential translating with constant velocity $v$, representing optical tweezers.
 A detailed description of the model is contained in Appendix A.}
 \label{fig:sim_schematics}
\end{figure}

\section{Model}
We introduce the mesoscopic model, aiming to capture some key elements of micellar networks and polymer solutions in a coarse-grained way (see Appendix A). 

An experimental probe is represented by a particle with effective radius $\sigma_0$, pulled by a harmonic trap that simulates optical tweezers with stiffness $\kappa$ moving at constant velocity $v=\abs{\vec{v}}$ along the $x$ axis. 
As in experiments, the probe is immersed in a complex fluid (see the sketch in Fig.~\ref{fig:sim_schematics}.
Inspired by the ``multi-blob'' representation~\cite{pierleoni2007soft}, we describe filaments (i.e., polymers or micellar tubules) as harmonic chains composed of $L$ units. Each effective unit encompasses many monomers on the molecular scale and repels other units via a soft Gaussian potential with characteristic free energy $\epsilon$ and effective size $\sigma$. The neighbors in a chain are held together by loose harmonic springs with stiffness $\kappa_\text{p}$.
 Notice that the model correctly reproduces the scaling behavior for the diffusion coefficient of semi-dilute polymers in the absence of the probe (see Appendix B).
Finally, the units-probe repulsion is modeled by another Gaussian potential 
whose characteristic energy scale $\epsilon_0$ is much larger than $\epsilon$. 
Units and probe obey overdamped Brownian dynamics in two dimensions, with temperature $T$ and friction coefficients $\gamma$ and $\gamma_0$, respectively.

Our simulations consider chains with a typical size comparable to or larger than the probe; this range is realistic if a chain represents a typical micellar tubule. 
These tubules may undergo a scission reaction paying a (relatively high) free energy cost $U$, resulting in a fluctuating length with average $ \avg{L_{t}} \simeq \phi^{1/2} \exp\{U/2 k_B T\}$~\cite{cates1990statics}, which gives $\avg{L_t} \approx 10^6$ for $U\approx 30 k_B T$ and a volume fraction $\phi=0.1$. 
For intermolecular distances $d \sim 1\,\text{nm}$, the mean countour length is $\avg{L_t} d \gg \sigma_0$, given that $\sigma_0\approx 3\,\mu\text{m}$.
Although the scission reaction in micellar fluids may be studied using models with patchy particles~\cite{baiesi2021rise,iubini2020aging}, it is not currently feasible to scale these latter models to nonequilibrium molecular dynamics simulations with a probe orders of magnitude larger than microscopic constituents.
Here, instead, we leverage the coarse graining and the topological constraint arising from the two-dimensional setting to emulate the scission reaction by the possibility that two units close to the probe are sufficiently separated by its repulsive force, allowing the probe to pass through.
We propose that this mechanism captures more complex reconfigurations that occur among entangled micellar tubules or polymers in three dimensions. 
In our model, these rearrangements release the elastic stress accumulated in front of the probe when it becomes comparable to the finite energy scale set by the Gaussian repulsive interactions.
This energy represents the dominant stress relaxation mechanism among many possible chemical or topological ones (scissions and recombinations~\cite{cates1990statics}, reptation~\cite{degennes1979scaling, doi1988theory}, etc.).

\begin{figure}
\centering
\includegraphics[width=\columnwidth]{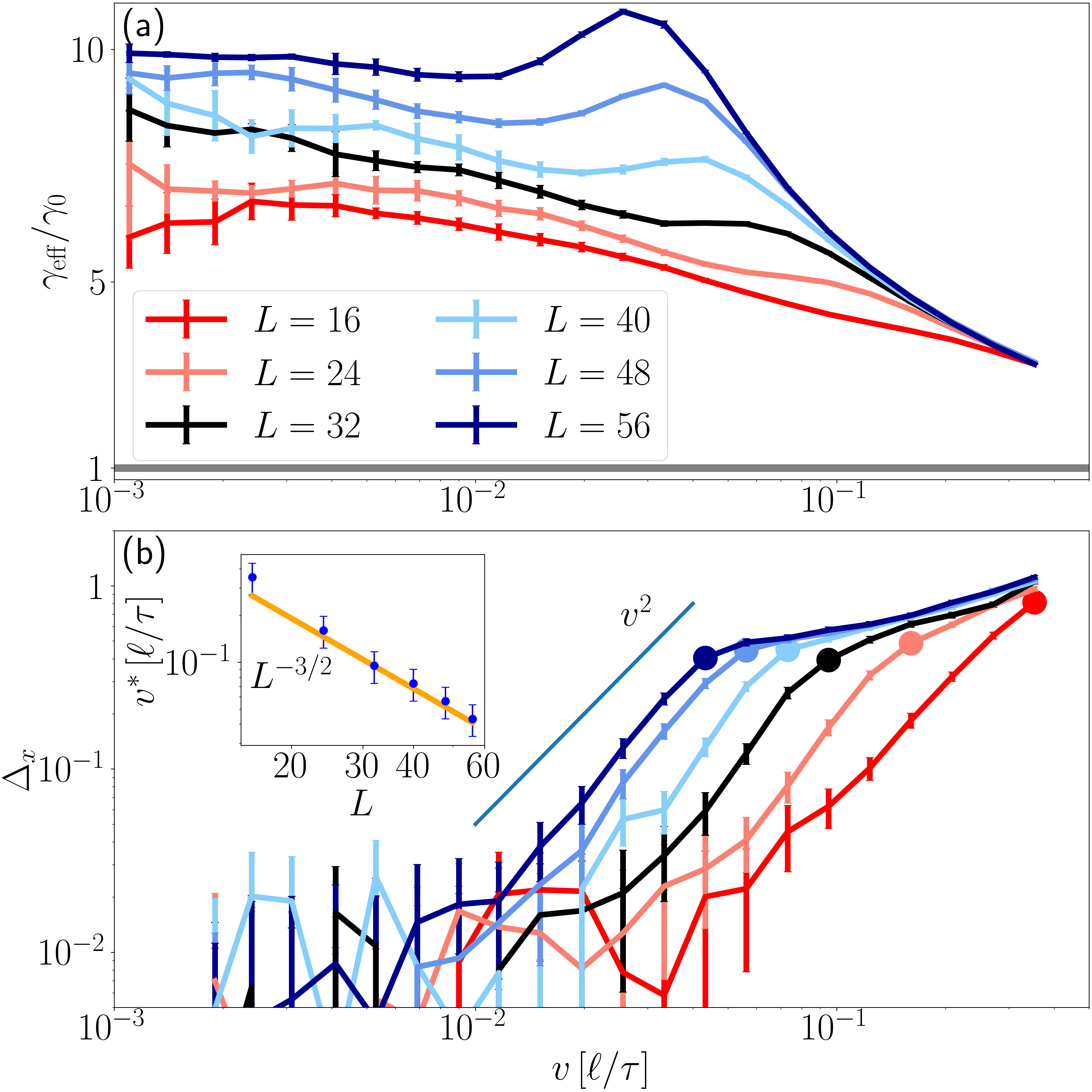}
 \caption{ (a) Semi-log plot of the effective friction $\gamma_{\text{eff}}$ (see Eq.~\eqref{eq:effective_friction_def}, in units of $\gamma_0$), as a function of the drag velocity $v$ (in units of $\ell/\tau$,  $\ell$ and $\tau$ being arbitrary simulation units for length and time resp.) for different polymer lengths $L$.
Error bars are obtained from the $r_{0,x}$ traces as the minimum between the naive estimate $\textrm{Var}\,r_{0,x}^{1/2} / v$ (uninformative for  $v\lesssim 10^{-2}$ due to near-equilibrium Brownian motion) and the one obtained from the linear fit $\langle r_{0,x}\rangle = a v $ on the points $(v_{n+k}, \langle r_{0,x}\rangle_{n+k})$ with $k=0, \pm 1, \pm 2$ (only informative in the near-equilibrium linear regime).
Gray line marks $\gamma_{\text{eff}}=\gamma_0$.
 (b) The same plot for the relative increase of the probe's variance from its equilibrium value; filled circles mark $v^*(L)$.
Error bars are computed via bootstrap, using $250$ sub-traces uniformly sampling $150$ different instantaneous values of the original $r_{0,x}$ traces.
 Inflection points (marked by circles) are identified by collapsing the data as in Fig.~\ref{fig:main}.
 Inset: Log-log plot of $v^*(L)$; the solid line shows the power law $\sim L^{-3/2}$. 
 Error bars are estimated as $\Delta v^* = v_n - v_{n-1}$ at $v_n=v^*$, where $\{v_n\}$ are the available values of the velocity.
 A linear fit (including uncertainty over $v$) of the function $\ln v = a\ln L + b$ gives $a=-1.51\pm 0.06$, $b=3.0\pm0.2$.
 }
 \label{fig:setup}
\end{figure}

\section{Results}
The first result of this work concerns the model's ability to replicate the multi-step friction curves~\cite{jain2021two} while enabling a detailed characterization of the fluid's microstructure and the probe's fluctuations. 

In Fig.~\ref{fig:setup}(a), for various polymer lengths $L$, we report the $v$-dependent effective friction coefficient
\begin{align}
 \gamma_{\text{eff}}(v) \equiv 
 \kappa \abs{\avg{r_{0,x}}}/v
 \,,\label{eq:effective_friction_def}
\end{align}
estimated from the average force exerted by the harmonic trap on the probe. 
In the following, $r_{0,x}$ denotes the probe coordinate on the $x$ axis in the trap frame.
For relatively short chains ($L=24,32$), there is a first plateau at small $v$ and a second at intermediate $v$.
Notably, the second plateau transitions into a shear-thickening peak for longer chains ($L>32$ in Fig~\ref{fig:setup}(a) for $10^{-2} \lesssim v \lesssim 10^{-1}$).
Other ways to model the multistep friction curves assume either that the probe is coupled to a few effective degrees of freedom diffusing in a rough potential landscape~\cite{jain2021two} or a temperature difference between the probe and the surrounding fluid~\cite{demery2019driven}.
Both approaches do not predict shear thickening for long filaments and, therefore, are experimentally discernible from the current one.

Crucially, we show that the variance $\text{Var}\, {r}_{0,x}$ of the probe's displacement from the trap center along the direction of the dragging display qualitatively similar curves, which are key to obtain mechanistic insight into this phenomenology.
In thermodynamic equilibrium, energy equipartition imposes $\text{Var}^{\text{eq}}\, {r}_{0,x} = k_BT/\kappa$ (see Appendix D). 
Thus, we quantify nonequilibrium effects in a steady state at constant $v$ by the relative fluctuations' enhancement 
\begin{align}
 \Delta_{x}(v) \equiv \frac{\text{Var}\, {r}_{0,x} -\text{Var}^{\text{eq}}\, {r}_{0,x} }{\text{Var}^{\text{eq}}\, {r}_{0,x} } = \frac{\kappa \text{Var}\, {r}_{0,x}}{k_B T} - 1 \,.
 \label{eq:Delta}
\end{align}
As shown in Fig.~\ref{fig:setup}(b), the departure of $\Delta_x(v)$ from zero is statistically negligible at low $v$. 
By increasing $v$, we observe the onset of the scaling $\Delta_x(v)\sim v^2$ before turning to a different behavior at the threshold velocity $v^*(L)$.
Hence, we introduce a dimensionless scale
\begin{align}
 \VV \equiv v / v^*(L)\,, \label{eq:v/v*}
\end{align}
that empirically distinguishes the regime $\VV\ll 1$, characterized by near-equilibrium linear response, from the nonlinear far-from-equilibrium regime at $ \VV \gtrsim 1$. 
The definition~\eqref{eq:v/v*} applies to any experimental system that displays the phenomenology of Fig.~\ref{fig:setup}(b).

\begin{figure}
\includegraphics[width=\columnwidth]{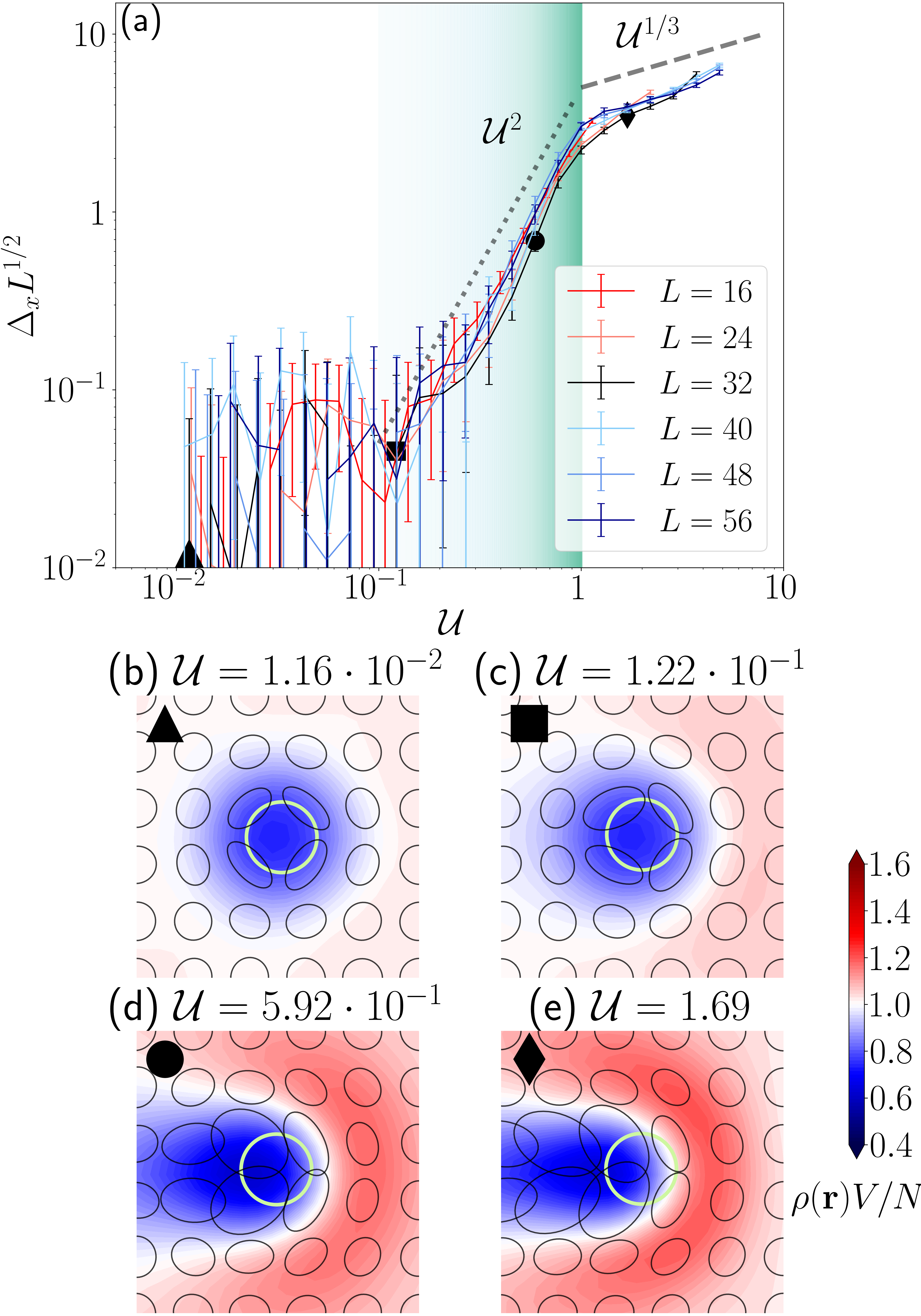}
\caption{(a) The fluctuations' enhancement~\eqref{eq:Delta} as a function of $\VV$, rescaled by $L^{1/2}$.
The green shaded region highlights the onset of the $v^2$ scaling and its end at $v=v^*$.
Error bars computed as in Fig.~2(b).
(b)-(e) Spatial fields of the local units density (in adimensional units, background color) and ellipses representing the gyration tensor $G$, for $L=32$ and increasing $\VV$, in a $20\ell\times 20\ell$ region around the probe (yellow-green circle of radius $2.5 \sigma_0$). 
Black symbols match those in panels (a) and help identify the regime of each panel. 
}
\label{fig:main}
\end{figure}

This leads to our central result. 
As shown in Fig.~\ref{fig:main}(a), the probe’s variance along the drag direction follows a master curve when plotted as a function of $\VV$, after a rescaling by $L^{1/2} $ (see Appendix D).
Notice that the existence of a master curve might be model-dependent. However, the initial growth of the fluctuations as $v^2$ is universal for symmetry reasons (explained later, see also Appendix D).
This implies that an analysis along the lines of the one that follows could be adapted also to experimental situations where a single master curve does not exist.

From the shape of $\Delta_x$ vs $\VV$, we can identify different dynamical regimes corresponding to structural changes in the polymer microstructure, quantified by the spatial field of the average gyration tensor $G(\vec{r}^\text{cm})=\langle\sum_{i,j=1}^L(\vec{r}_i-\vec{r}^{\text{cm}})(\vec{r}_j-\vec{r}^{\text{cm}})/L \rangle$, where $\vec{r}_i$ is the position of the unit $i$ and $\vec{r}^{\text{cm}}$ is the center of mass of the polymer. 
In Fig.~\ref{fig:main}(b)-(e), we show, for $L=32$, how $G$ (represented by ellipses) changes in the different regimes identified by the probe's fluctuations. 
These changes occur in parallel with the typical decrease in the density of local units (the background shade in Fig.~\ref{fig:main}(b)-(e)) in the wake of the probe, as in the case of colloidal suspensions~\cite{squires2005simple, zia2018active}.
In the low $\VV$ regime, which is due to a finite experimental sensitivity, the probe's nonequilibrium fluctuations are indistinguishable from zero within numerical errors and the profiles of the gyration tensor are qualitatively analogous to equilibrium ones, see Fig.~\ref{fig:main}(b), exhibiting an approximate spherical symmetry around the probe.
Upon increasing $\VV$, the variance grows as $\VV^2$, which corresponds to
a scaling $\sim v^2$ (see Appendix D). 
Correspondingly, the chains deform and the gyration tensor profiles develop anisotropies due to the flow. 
Chains experience significant stretching ahead of the probe and in its trailing wake, aligning with the flow, see Fig.~\ref{fig:main}(c)-(d).
Eventually, at large values of $\VV$, the probe's variance transitions to a different power law, while the average size of the chains in front of the probe becomes smaller, see Fig.~\ref{fig:main}(e).

\section{Discussion}

We now show how to link the observed phenomenology to the microscopic dynamics and characteristic length scales of our model. 
However, we expect that the main elements of this approach are generic and applicable to other complex fluids, where the underlying microscopic dynamics may differ.
In our model, we observe the scaling $v^*(L)\sim L^{-c}$ with $c\simeq 3/2$ (see inset of Fig.~\ref{fig:setup}(b)). 
In what follows, we identify two distinct mechanisms that explain the two crossovers between regimes at low $\VV$ and at $\VV \approx 1$, both entailing the observed scaling in $L$. 

\begin{figure*}
 \centering
 \includegraphics[width=\textwidth]{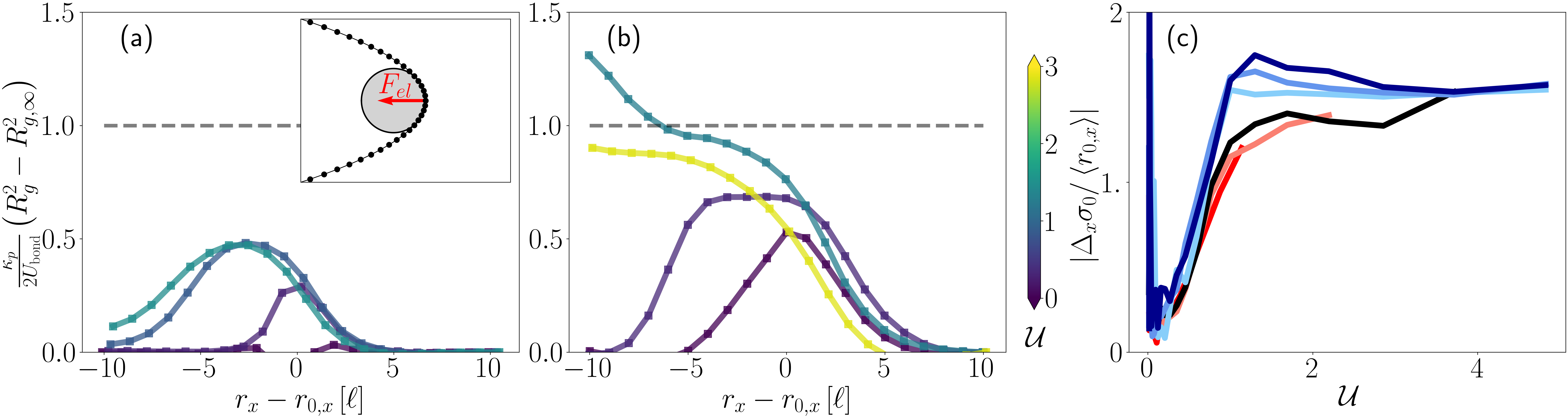}
\caption{
(a-b) Local elastic energy (in adimensional units) stored in the polymer fluid as a function of the distance from the probe along the $x$-axis, for (a) $L=16$ and (b) $L=56$.
Each curve is colored according to the value of $\VV$.
The dashed line represents the threshold energy $U_\text{bond}$ for the activated jump events.
In panel (a), missing points indicate the absence of polymers in that position (measured using the center of mass) for low $\mathcal{U}$, due to the repulsive force exerted by the probe. 
$R^2_{g,\infty}$ is the value of the gyration radius computed far away from the probe.
Inset: a sketch of the configurations giving rise to hoop stresses.
(c) Dimensionless ratio of the fluctuations' enhancement to the mean probe position scaled by $\sigma_0$, for various $L$ (color code in Fig.~\ref{fig:main}). For $\VV >1$, the ratio tends to a constant value independent of $L$. 
}
\label{fig:radius_of_gyration_profile}
\end{figure*}

We first deal with the first crossover, from equilibrium to the $\sim \VV^2$ scaling at $\VV\approx 10^{-1}$. 
Near equilibrium, our model for linear chains with negligible entanglement is characterized by the competition of single-polymer diffusion and advection. 
The diffusion coefficient of the chains, to leading order in $L$, has the same scaling as the Rouse model~\cite{doi1988theory}, $D_L=k_BT/\gamma L$ (see Appendix B). 
The resulting effective velocity associated with diffusion is $v^P(L)\equiv D_L/R_g^\text{eq}$, where the gyration radius $R_g^\text{eq}$ can be approximated using the Rouse model, namely $R_g^\text{eq}=\sqrt{L k_BT/6\kappa_\text{p}}$. 
Therefore, the near-equilibrium dynamics of the system is characterized solely by the ratio of the advection velocity $v$ and the diffusion velocity $v^P(L)$, that is, the P\'eclet number~\cite{van2024soft}
\begin{align}\label{eq:Peclet}
\text{Pe} \equiv \frac{v }{v^P} =\frac{v R_g^{\text{eq}}}{D_L} = \frac{\gamma v L^{3/2}}{\sqrt{6\kappa_{\mathrm{p}} k_BT}}\,.
\end{align}
Since $\text{Pe} \sim \VV \sim L^{3/2} $, the analysis explains the collapse reported in Fig.~\ref{fig:setup}(a) in the regime $\VV \lesssim 1$. 
However, the data collapse continues outside of the linear response regime $\VV \lesssim 1$. 
Furthermore, the P\'eclet number determined by computing $D_L$ and $R_g$ numerically is $\text{Pe} \approx \VV/10$ (see Appendix B).
Hence, by increasing $\VV$, at $\VV \approx 10^{-1}$ we observe the crossover to advection-dominated dynamics, \emph{i.e.}, $\text{Pe}\approx 1$, where chains do not have enough time to diffuse freely away from the approaching probe and suffer substantial conformational changes leading to stress accumulation.
The quantitative difference between $\VV$ and $\text{Pe}$ suggests that a different kinetic mechanism is at work far from equilibrium.
We note that the commonly used Weissenberg number~\cite{furst2017microrheology, chapman2014nonlinear, jain2021two} $\textrm{Wi} \equiv v \tau_f/\sigma_0$, where $\tau_f$ is a relaxation time of the fluid, would have a different scaling from the one we find for $\VV$ and $\text{Pe}$, because the typical size used to define it is the probe size rather than the size of a polymer.

After the first crossover, the fluctuation enhancement scales as $\mathcal{U}^2$, similarly to what is found for a probe driven in a Gaussian field~\cite{demery2019driven}.
This results from symmetry considerations, which we briefly mention here and develop further in Appendix D.
In thermodynamic equilibrium $\text{Var}^\text{eq}\,\vec{r}_0 = k_BT/\kappa$, corresponding to a null fluctuation enhancement, $\Delta_x=0$ due the translational invariance of the canonical equilibrium distribution (conditioned on the probe position).
Moreover, $\Delta_x=O(v^2)$ for small drag speed, as the variance is left unchanged by switching to a reference frame where $\vec{v}'=-\vec{v}$, ruling out a linear term in the expansion.

To understand the second crossover, in $\Delta_x$ at $\VV \approx 1$, and to link it with the nonlinear behavior of the friction curves, we study the increase in the average local elastic energy of the chains, detected by the difference between $R_g^2$ 
at a given $\VV$ and the equilibrium one (at $\VV=0$).
Fig.~\ref{fig:radius_of_gyration_profile}(a)-(b) shows such quantity as a function of the distance from the probe along the $x$ axis for different values of $\VV$.
For all $L$ in our simulations, the chains increase their elastic energy at large $\VV$ as the probe deforms them into hook-shaped configurations (inset of Fig.~\ref{fig:radius_of_gyration_profile}(a)), until they have acquired enough energy to escape.
The theory gives a threshold energy $U_\mathrm{bond}$ (derived later, see also Appendix E), used to rescale the energy in Fig.~\ref{fig:radius_of_gyration_profile}(a)-(b). 
Short chains (Fig.~\ref{fig:radius_of_gyration_profile}(a)) accumulate little elastic energy by increasing $\VV$, and therefore rarely escape from the probe. 
Long chains, instead, deform sizably and experience an increased effective friction due to the resulting {hoop stresses} \cite{van2024soft}, until an escape event is triggered.
Starting from $L=24$, the elastic energy becomes comparable to $U_\mathrm{bond}$, causing a polymer to escape primarily through a single bond ``scission'', which is an overextension of a single polymer bond that allows passage of the probe.
This provides a mechanism for local stress relaxation in front of the probe, particularly noticeable at high $\VV$ values, as shown in Fig.~\ref{fig:radius_of_gyration_profile}(b) for the yellow curve corresponding to $\mathcal{U}\simeq 3$.

Such structural changes at large $\VV$ mark a dramatic variation in the statistics of the probe position. As shown in Fig.~\ref{fig:radius_of_gyration_profile}(c), the fluctuations' enhancement $\Delta_x$ becomes proportional to the mean probe displacement $\avg{r_{0,x}}$ at $\VV \gtrsim 1$.
It signals that the probe's dynamics becomes dominated by Markovian jumps, which we identify with discrete activated events in which a stretched polymer suddenly releases the accumulated stress. 
A simple model for probe displacement can be derived by coarse-graining the full Langevin dynamics in the low-density limit (see Appendix E), which predicts the asymptotic plateau shown in Fig.~\ref{fig:radius_of_gyration_profile}(c). Moreover, it identifies $v^* \sim L^{-3/2} \sqrt{U_\text{bond}}$ as the threshold velocity at which the accumulated elastic energy $U_\text{el}$ equals the energy $U_\text{bond}$ required to stretch a polymer bond. The elastic energy $U_\text{el} \sim v^2 L^3$ stored in the polymer is obtained from the mean equilibrium energy of a Gaussian chain of size $\sim L$ and with a pinned end, subject to a constant force field of magnitude $\gamma v$. The energy $U_\text{bond}$ is estimated by selecting the single bond distance $2y^*$ that minimizes the sum of attractive energy between two units and repulsive energy between the two units and the probe. 
In our simulations, stretching a bond to the distance $ 2y^* \approx 4\sigma_0$, needed for the passage of the probe through it, requires $U_\text{bond}\approx 30 k_BT$, compatible with the scission free energy cost $U$, introduced previously~\cite{cates1990statics, jain2021two}.
Hence, our model mimics the configurational rearrangement that the traveling probe induces in experiments with living polymers \cite{gomez2015transient, berner2018oscillating, jain2021two}.

\begin{figure*}
    \centering
    \includegraphics[width=\textwidth]{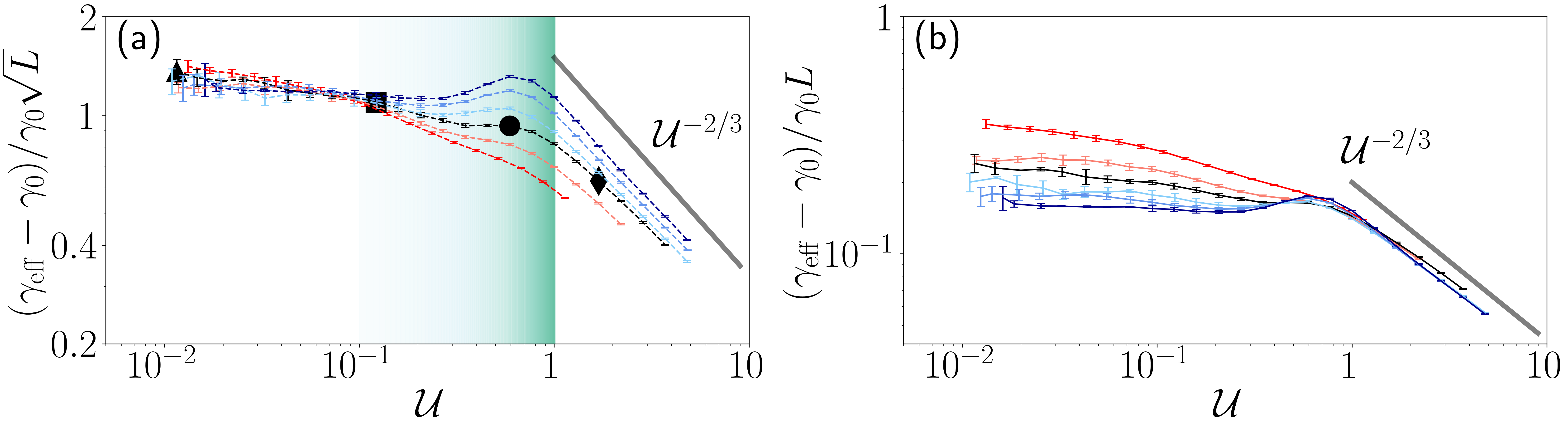}
    \caption{(a) Difference between the probe's effective friction and its Stokes friction $\gamma_0$, in units of $\gamma_0$ and rescaled by $\sqrt{L}$. 
    Symbols and color code for $L$ are the same of Fig.~3. 
    (b) As in panel (a) but rescaled by $L$.}
\label{fig:scaled_effective_friction}
\end{figure*}

The microstructural regimes identified above directly impact the average position of the probe, determining the value of the effective friction measured. 
Figure~\ref{fig:scaled_effective_friction}(a)-(b)
illustrates how the effective friction~\eqref{eq:effective_friction_def} changes relative to $\VV$, and the different scaling in $L$.
As detailed in Appendix A, the chains cause an additional average viscous friction force on the probe, proportional to $\gamma_{\text{eff}} - \gamma_0 \sim \gamma L \sigma_0 / R_g$, which, in steady state, is compensated by the force exerted by the tweezers.
As a result, in the linear response regime, when $R_g \sim \sqrt{L}$, the excess friction scales like $\sqrt{L}$.
Out of equilibrium, it is necessary to consider the polymer stretching $R_g\sim L$, observed at $\VV>1$.
Assuming that $R_g$ depends on the velocity through a power law in $\mathcal{U}$, i.e., $R_g=R_g(\mathcal{U})\sim \mathcal{U}^\alpha$, we obtain $\alpha=2/3$ and 
\begin{align}
\gamma_{\text{eff}} -\gamma_0 \sim L \VV^{-2/3}\,,
\label{eq:scaling_friction_high_Mi}
\end{align}
as confirmed by the data collapse in the inset of Fig.~\ref{fig:radius_of_gyration_profile}(c). 
Furthermore, due to the proportionality $\Delta_x \sim \langle r_{0,x}\rangle \sim \gamma_{\text{eff}} v \sim L^{-1/2}\mathcal{U}^{1/3}$ in the regime $\VV >1$ (see Fig.~\ref{fig:radius_of_gyration_profile}(c) and Appendix E), we also explain the second scaling $\Delta_x~\sim \VV^{1/3}$ observed in Fig.~\ref{fig:main}(a) for $\VV >1$.

Note that thickening and multi-step friction curves are distinct quantitative manifestations of the same fundamental mechanism---the localized accumulation of elastic energy near the probe, which generates hoop stresses~\cite{van2024soft}---made possible by the repulsive energy scale imposed by the probe itself. Our model suggests that this mechanism is present not only in slab-confined polymeric fluids but also (when viewed as an effective low-dimensional coarse-grained system) in three-dimensional complex fluids capable of sustaining substantial local stress buildup associated with transient kinetic trapping of the constituents. Therefore, the non-continuum nature of this effect likely explains why complex fluids with multi-step friction curves in microrheology experiments exhibit monotonic flow curves in bulk rheology~\cite{jain2021two}.

\section{Conclusions}
Analyzing the variance of a probe forced through a complex medium (i) yields an adimensional scale for nonequilibrium based on the breakdown of equipartition, (ii) reveals the occurrence of discrete stress releases in an otherwise diffusive system, and (iii) allows identifying the proper timescales of the fluid's microstructural deformations. 
Determining the relevant timescales of complex fluids from fluctuations of microprobes is agnostic of the specific system and should help analyze experimental data, e.g., for complex fluids as micellar networks~\cite{berner2018oscillating, jain2021two}. 
Hence, our suggestion to combine the analysis of nonequilibrium fluctuations of the probe with the effective friction curves may help formulating mechanistic explanations of both experimental and theoretical findings on complex and active media.

Our coarse-grained model reproduces the phenomenology found in active microrheology experiments, and some preliminary data on recoil measurements show two well-separated relaxation time scales as found in experiments~\cite{gomez2015transient, ginot2022recoil}.
Some of the other features remain to be explored.
Since our focus here is on the noncontinuum effects when the probe size is comparable to that of the chain, we did not systematically explore the parameter space of the model in other situations. 
In particular, we expect a crossover to the macroscopic situation when the probe size $\sigma_0$ is increased by orders of magnitude.
In the future, we will pursue some of these directions.

Practical applications and extensions of our approach may include determining the effective size of the mesoscopic objects deformed by the active probe~\cite{yang2023local}, linking the recently found spatial profiles of dissipation~\cite{venturelli2024stochastic} to the localized elastic stresses that we unveil in this study, or ascertaining the role of local elastic stresses in the mechanical characterization of active fluids~\cite{paul2022force, conforto2024fluidification}.

\vspace{2mm}
\begin{acknowledgments}
The authors thank Enrico Carlon and Sabine Klapp for valuable discussions. This work is supported by the projects ``BAIE\_BIRD2021\_01'' and ``FALASCO PARD 2023'' of the University of Padova and by the MIUR grant Rita Levi Montalcini. 
The authors acknowledge the use of CloudVeneto for computing and data storage.
\end{acknowledgments}

\appendix

\section{Details on the molecular dynamics simulations}

\begin{figure*}
\centering
\includegraphics[width=0.99\textwidth]{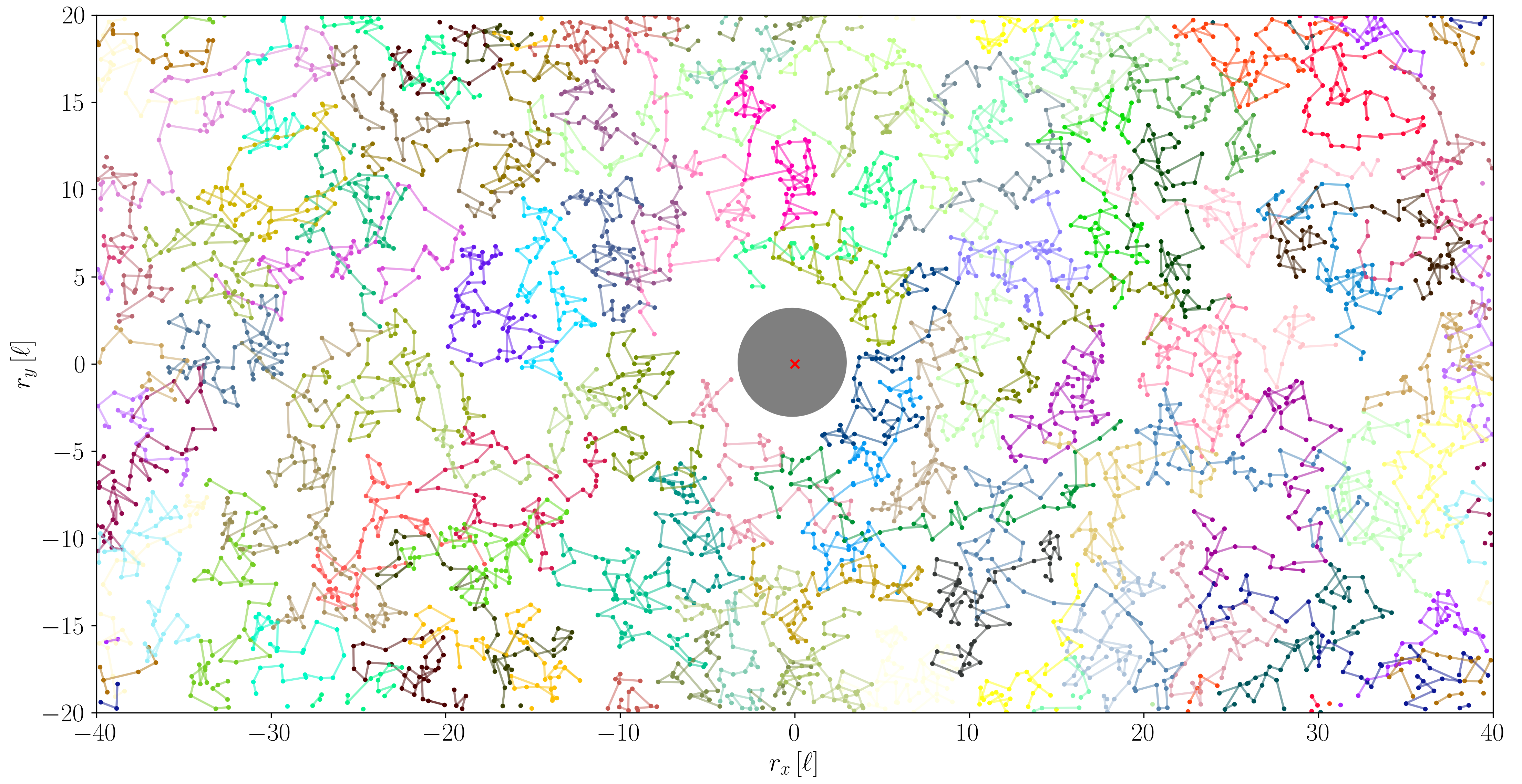}
\caption{Snapshot of a simulation with $L=56$, $v=1.97\times 10^{-2} \ell/\tau$, box size $80\ell \times 40\ell$, and $4704$ units (yielding a number density $\rho\simeq 1.5$ because of the volume excluded by the probe). 
Each chain is shown in a different color. 
The red cross is the center of the harmonic trap, and the grey disk has a radius $2.5\sigma_0$ around the position $\mathbf{r}_0$ of the probe.}
\label{fig:snapshot}
\end{figure*}

The numerical model comprises $N$ units, indexed with $1\le i \le N$, and a probe particle, indexed with $i=0$. 
Both species are modeled as Brownian particles in contact with a bath at temperature $T$. 
Each unit represents a coarse-grained portion of a polymer, characterized by friction coefficient $\gamma$ and diffusion constant $D = k_B T/\gamma$. 
Units interact with each other via a soft repulsive Gaussian potential $U(r)=\epsilon\, e^{-r^2/2\sigma^2}$ with an energy scale $\epsilon = k_B T$. 
We arrange the units in $M=N/L$ linear chains with $L$ units each, and $L-1$ springs with elastic constant $\kappa_{\mathrm{p}}$ join neighboring units. 
Fig.~\ref{fig:snapshot} shows a snapshot from a simulation ($L=56$, $v = 0.0197 \ell/\tau$).

The probe's Brownian motion is characterized by $\gamma_0>\gamma$ and $D_0=k_BT/\gamma_0 < D$ as befits a larger solid particle. 
It interacts with the units also via the Gaussian potential $U_0(r)=\epsilon_0 e^{-r^2/2 \sigma_0^2}$, with an energy scale $\epsilon_0\gg k_B T$ that ensures little overlap between the chains and the probe.

The positions of the particles in the solvent frame are denoted by $\vec{q}_i$. We perform the simulations in a two-dimensional box of size $\mathcal{B}_x \times \mathcal{B}_y$ with periodic boundary conditions.
The box size is at least $\mathcal{V} = 80\ell \times 40\ell$.

The system also contains optical tweezers that move at constant velocity $v=|\vec{v}|$, with $\vec{v}$ aligned with the $x$ axis. 
The overdamped Langevin dynamics for the coordinates in the trap's co-moving frame, $\vec{r}_i = \vec{q}_i-\vec{v}t$, reads 
\begin{subequations}
\label{eq:langevin}
\begin{align}
    & \dot{\vec{r}}_0 =  -\vec{v} + \gamma_0^{-1}\left[ \vec{F}_{\text{tw}} +\sum_{i=1}^N \vec{F}_0(\vec{r}_{0i}) \right]+\sqrt{2 D_0}\vec{\xi}_0\,, \label{eq:probe_comoving_langevin}\\
    & \dot{\vec{r}}_i =  -\vec{v} + \gamma^{-1}\left[\vec{F}_0(\vec{r}_{i0}) +\sum_{j=1}^N \vec{F}^{\text{int}}(\vec{r}_{ij})\right]
    +\sqrt{2 D}  \vec{\xi}_i \,,\label{eq:monomer_comoving_langevin}
\end{align}
\end{subequations}
where $\vec{r}_{ij} = \vec{r}_i - \vec{r}_j$ and each $\vec{\xi}_i$ is an independent white noise.
In the drifts, $ \vec{F}_{\text{tw}} = - \nabla U_{\text{tw}}= - \kappa\vec{r}_0$ is the force exerted by the tweezers, $\vec{F}_0 = - \nabla U_0$ is the repulsive force between the probe and each unit and $\vec{F}^\text{int}$ contains both the units repulsive interaction $\vec{F}=-\nabla U$ and the elastic attraction inside the $m$-th chain, $\kappa\sum_{j=1}^N I_{i,m}I_{j,m}(\delta_{j,i+1}+\delta_{j,i-1})\left(\vec{r}_j-\vec{r}_i \right)$, where $I_{i,m}$ is the indicator function of the $i$-th unit in the $m$-th chain. 
The reciprocity for the force between one unit and the probe reads explicitly $\vec{F}_0(\vec{r}_{0i}) = - \vec{F}_0(\vec{r}_{i0})$. 
We solve the equations of motion in the solvent frame, employing the Brownian module of LAMMPS~\cite{thompson2022lammps}.
The values of all parameters are reported in Tab.~\ref{tab:parameters}. 
After reaching the steady state for each set of values of $L$ and $v$, we run simulations for at least $5 \cdot 10^5 \tau$ and run at least $4$ independent realizations to determine the average displacement of the probe from the trap and the corresponding variance.

\begin{table}[t!b]
    \centering
    \begin{tabular}{l|l|l}
\hline  
        \textbf{Physical quantity} & \textbf{Symbol} & \textbf{ Value}\\[1.5pt]
        \hline
        Friction coefficient of the probe & $\gamma_0$ & $20\, \mathcal{E} \tau/\ell^2$ \\
        Friction coefficient of a particle & $\gamma$ & $ \mathcal{E}\tau/\ell^2$ \\
        Probe-particle interaction distance & $\sigma_0$ & $11/8\,\ell$\\
        Particle-particle interaction distance & $\sigma$ & $1/4\,\ell$\\
        Probe-particle max energy & $\epsilon_0$ & $10\,  \mathcal{E}$ \\
        Particle-particle max energy & $\epsilon$ & $1/2\,\mathcal{E}$ \\
        Thermal energy & $k_B T$ & $1/2\,\mathcal{E}$ \\
        Stiffness of the trap & $\kappa$ & $25\, \mathcal{E} / \ell^2$ \\
        Stiffness of the chain bond & $\kappa_{\mathrm{p}}$ & $\mathcal{E}/ \ell^2$\\
        Number density & $\rho_0=N/\mathcal{V}$ & $1.5\,\ell^{-2}$ \\
        Numerical time step & $dt$ & $5 \cdot 10^{-3}\tau$
        \\[1.5pt]
        \hline
    \end{tabular}
    \caption{{Parameters of the model}. 
    We set $\ell$, $\tau$, and $\mathcal{E}$ as the units of length, time and energy, respectively. }
    \label{tab:parameters}
\end{table}

\vspace{3mm}
\section{Polymer dynamics}

We discuss the conformation and dynamics of the chains in simulations without the probe (formally, obtained setting $\epsilon_0=0$ in Eqs.~\eqref{eq:langevin}).

\begin{figure*}
    \centering
    \includegraphics[width=\textwidth]{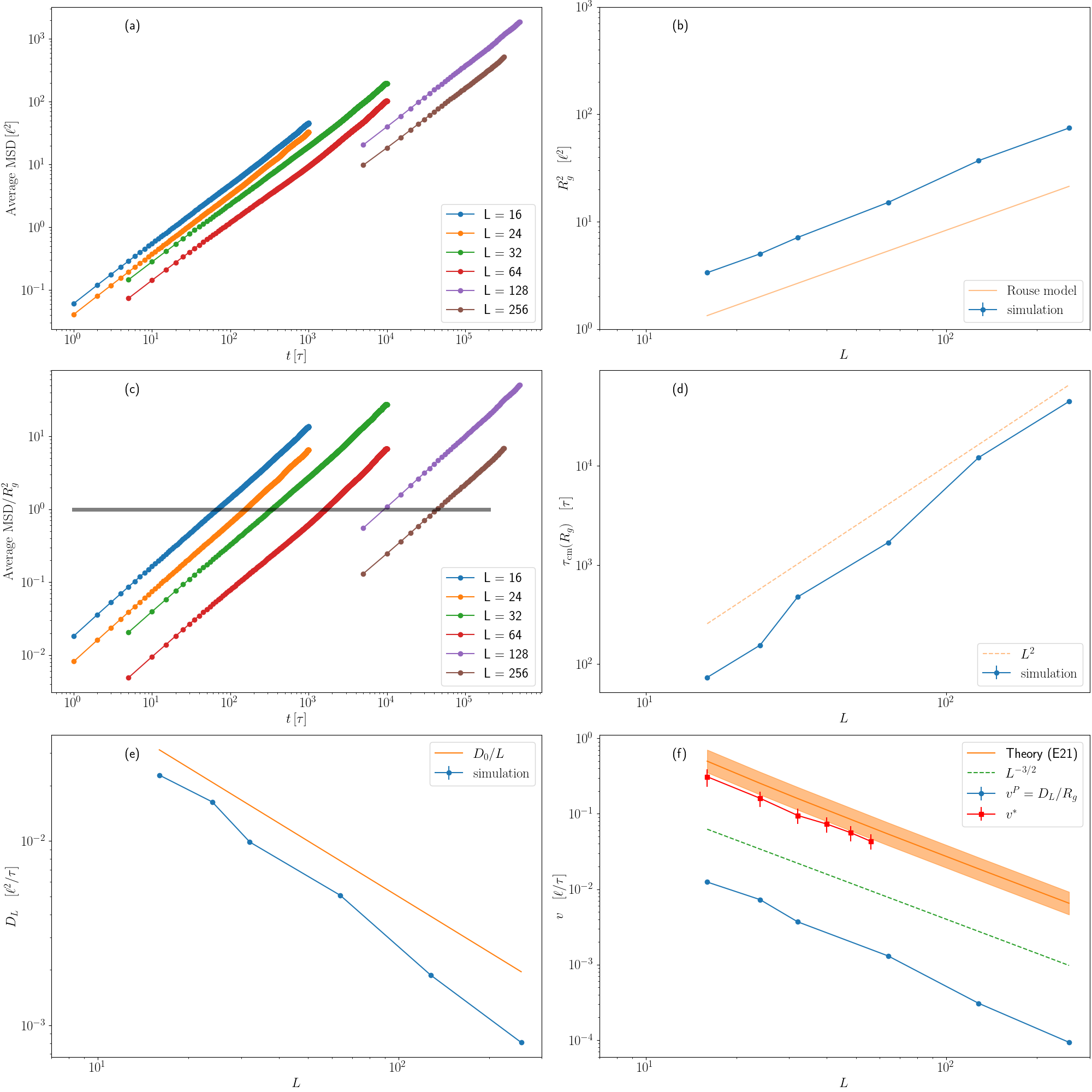}
    \caption{Free diffusion of the chains, with number density $\rho_0=1.5 \ell^{-2}$. In all panels, invisible error bars are smaller than the symbols' size. (a) Single-chain mean-squared displacement as a function of time, varying the chain length $L$.
    (b) Mean-squared radius of gyration as a function of $L$.  
    (c) Mean-squared displacements, divided by $R_g^2(L)$. The crossing with the horizontal line identifies the diffusive relaxation time $\tau_{\textrm{cm}}(R_g)$.
    (d) Diffusive relaxation time of the chains extracted from the crossings in panel (c).
    Error bars (extracted from linear interpolation) are comparable to the symbol size.
    (e) Diffusion coefficients of the chains, extracted from (a) with a linear fit. 
    Error bars are comparable to the symbol size.
    (f) Comparison between the critical velocity scale derived from diffusion, $v^P$, the one found empirically from the probe's fluctuations $v^*$, and the theoretical prediction in \eqref{eq:vstartheory}.
    The shaded stripe shows the sensitivity of the theoretical prediction in Eq.~\eqref{eq:vstartheory} to the uncertainty in the threshold energy $U_\mathrm{bond}$,
    obtained in Eq.~\eqref{eq:U} for a drastically oversimplified situation (considering only a single pair of units).
    The upper limit corresponds to $U_\mathrm{bond}=2U(y^*)$, the center line to $U_\mathrm{bond}=U(y^*)$ and the lower one to $U_\mathrm{bond}=U(y^*)/2$.}
    \label{fig:msd_and_diffusion}
\end{figure*}

First, we discuss the limiting case $L=1$ (simulations not shown), corresponding to the Gaussian core model~\cite{stillinger1976phase}, which is known to have a glass phase for $k_BT\ll \epsilon$ \cite{lang2000fluid, ikeda2011glass, prestipino2005phase}.
Our parameters, in particular the choice $ k_BT  = \epsilon $ for the repulsive Gaussian potential for the interaction between two units, guarantee that the simulations with $L=1$ remain within the liquid phase irrespective of the density, as determined from the phase diagram in Ref.~\cite{prestipino2005phase}.

Second, we study the diffusion of the chains' center of mass, compared to the theoretical estimates derived from the Rouse model of polymers~\cite{doi1988theory}.
The mean-squared displacements for chains of different lengths $L$ follow a diffusive scaling at all times, as shown in Fig.~\ref{fig:msd_and_diffusion}(a).
The mean-squared radius of gyration and the diffusion coefficients are compatible (within a factor $\approx 3$) with the predictions of the Rouse model, as reported in Fig.~\ref{fig:msd_and_diffusion}(b) and (e). 
This confirms that our simulations are performed in the semidilute regime, where the mean radius of gyration scales like $R_g\sim L^{\nu}$, with $\nu=1/2$;  the repulsive potential between units, which would otherwise modify the exponent to approximatively $\nu=3/4$ (self-avoiding walk in 2D), is effectively masked \cite{vanderzande1998lattice}.

The scaling of the relaxation time for the diffusion of the center of mass, estimated using the time at which the mean-squared displacement reaches the radius of gyration, is shown in Fig.~\ref{fig:msd_and_diffusion}(d), and aligns reasonably well with the one of a Gaussian polymer, $\tau_{\textrm{cm}}(R_g) \sim L^{2\nu+1}=L^2$~\cite{doi1988theory}.

Finally, in Fig.~\ref{fig:msd_and_diffusion}(e), we show the typical scale $v^P = D_L/R_g$ for the velocity of the center of mass diffusion, confirming the scaling $L^{-3/2}$ predicted with the Rouse model and compare it with $v^*$ obtained from simulations, following the procedure explained in the main text, and with the theoretical prediction $v_\text{th}^*$ obtained in Sec.~\ref{sec:hopping}. 


\section{Scaling of the effective friction at low and high $\mathcal{U}$}

In the steady state, with averages denoted by $\avg{\ldots}$,  the local average particle density reads
  $ \rho(\vec{r}) \equiv \frac{1}{N} \sum_{i=1}^N \avg{\delta(\vec{r}-\vec{r}_i)}$
and the velocity field 
$ \vec{V}(\vec{r}) \equiv {\rho(\vec{r})}^{-1} \frac{1}{N} \sum_{i=1}^N (\avg{\dot{\vec{r}}_i+\vec{v})\delta(\vec{r}-\vec{r}_i)}$,
which equals $\vec{v}$ only if the units around $\vec r$ are at rest in the co-moving field.
Since $\avg{\dot{\vec{r}}_0}=0$, the average force of the tweezers balances out the total friction on the probe. 
Combining Eqs.~\eqref{eq:probe_comoving_langevin}, \eqref{eq:monomer_comoving_langevin}, and 
using internal forces' reciprocity leads to 
\begin{align}
\begin{split}
    \avg{\vec{F}_{\text{tw}}}    &= \gamma_0\vec{v} +N\gamma \int \d\vec{r} {\rho(\vec{r})} {\vec{V}(\vec{r})}\,. \label{eq:tweezers_force}
\end{split}
\end{align}
whose modulus, divided by $v$, gives the effective friction
\begin{align}
    \gamma_{\text{eff}}(v)&\equiv \frac{\abs{\avg{\vec{F}_{\text{tw}}} }}{v} = \gamma_0 + N \gamma  \int \d\vec{r}\, \rho(\vec{r})\frac{{\vec{V}(\vec{r})}\cdot\vec{v}}{v^2} \,.\label{eq:effective_friction}
\end{align}

We can extract useful qualitative information through scaling arguments for the friction coefficient. First, we define the local density from the $m$-th chain, $\rho_m(\vec{r})=\sum_{i=1}^N I_{i,m} \delta(\vec{r}_i-\vec{r})/L$, where $I_{i,m}=1$ if the $i$-th unit belongs to the $m$-th chain, and $I_{i,m}=0$ otherwise. 
We then rewrite \eqref{eq:tweezers_force} as 
\begin{align}
    \avg{\vec{F}_{\text{tw}}}  = \gamma_0\vec{v} +L\gamma \sum_{m} \int \d\vec{r} \rho_m(\vec{r})\vec{V}(\vec{r})\,.
\end{align}
Sufficiently far from the probe particle, the motion of a chain is unperturbed, i.e., ${\vec{V}(\vec{r})}=\vec{0}$, while close to the probe, an approximate no-slip condition must hold, ${\vec{V}(\vec{r})}\approx\vec{v}$. 
Because of this, the sum over $m$ contains only a small number of contributions from the chains close to the probe, which for a large enough $L$ will scale like $O(L^0)$ (as it cannot decrease below $1$). 
Furthermore, if the chains are long enough, the number of units to which the no-slip condition applies can be estimated by the participation ratio $\sigma_0/R_g$.
Since $R_g \sim \sqrt{L}$ for all chains at low $v$, we get 
\begin{align}
    \avg{\vec{F}_{\text{tw}}} - \gamma_0\vec{v} \approx L\gamma \frac{\sigma_0}{R_g} \vec{v} \approx  \gamma \sqrt{L} \vec{v}\label{eq:scaling_friction_low_v}\,,
\end{align}
which explains the plateau for small $\mathcal{U}$ in  Fig.~\ref{fig:scaled_effective_friction}(a).

At high $v$, the probe stretches the chains in its proximity, whose typical size is expected to scale linearly with $L$~\cite{degennes1979scaling}. 
Assuming that the typical size of the chains depends on the drag velocity only via the adimensional velocity $\mathcal{U}$, and requiring the expected linear scaling in $R_{g}\sim L$ leads to
\begin{align}
    R_g(\mathcal{U}) \sim \mathcal{U}^{2/3} \sim v^{2/3}L\,. \label{eq:high_pe_radius_gyration}
\end{align}
(We recall that $\mathcal{U}$ is defined in Eq.~(3) of the main text, and that the analysis in  Sec. V  of the SM yields $\mathcal{U}\sim vL^{3/2}$, confirming the near-equilibrium scaling of Eq.~(4) of the main text.
Using \eqref{eq:high_pe_radius_gyration} to estimate the participation ratio in \eqref{eq:scaling_friction_low_v}, now extended to the high $\mathcal{U}$ case, gives
\begin{align}
    \gamma_\text{eff}(v) -\gamma_0 = \frac{ \avg{\vec{F}_{\text{tw}}} }{v} -\gamma_0 \sim  \gamma L \mathcal{U}^{-2/3}  \sim \gamma v^{-2/3}\,.
    \label{eq:2/3}
\end{align}
This is the scaling behavior shown in Fig.~3(c) of the main text (see also Fig.~\ref{fig:scaled_effective_friction}(b)).

\section{Variance of the probe} 

We study the variance $\text{Var}\,\vec{r}_0\equiv \avg{\vec{r}_0^2}-\avg{\vec{r}_0}^2$ of the probe's position relative to the trap's center.
Multiplying \eqref{eq:probe_comoving_langevin} by $\vec{r}_0$ with the Stratonovich product~\cite{van2024soft} (denoted by $\circ$) to obtain a stationary state average, $\d_t\avg{\vec{r}_0^2}=0$, on the left-hand side, we obtain
\begin{align}
    \begin{split}
    \kappa\avg{\vec{r}_0^2} &=  -\gamma_0\vec{v}\cdot\avg{\vec{r}_0} + \sum_i \avg{\vec{r}_0\cdot\vec{F}_0(\vec{r}_{0i})} \\
    &\quad+ \sqrt{2\gamma_0k_BT }\avg{\vec{r}_0(t)\circ\vec{\xi}_0(t)}\,.
\end{split}
\end{align}

 Then, we use \eqref{eq:probe_comoving_langevin} to solve for $\gamma_0\vec{v}$.
We have that $\avg{\vec{r}_0(t)\circ\vec{\xi}_0(t)}=\sqrt{\gamma_0k_BT/2}$, $\avg{\vec{r}_0(t)\circ\vec{\xi}_i(t)}=0$ for every unit $i$ and $\avg{\dot{\vec{r}}_0}=0$ in a steady state. Thus, we arrive at the following expression involving $\text{Var}\,\vec{r}_0$,
\begin{align}
\begin{split}
     \text{Var}\,\vec{r}_0 
     &= \frac{k_BT}{\kappa}  + \frac{1}{\kappa} \sum_i \big( \avg{\vec{r}_0\cdot \vec{F}_0(\vec{r}_{0i})} - \avg{\vec{r}_0}\cdot\avg{\vec{F}_0(\vec{r}_{0i})}\big)\,. \label{eq:variance_r0}
\end{split}
\end{align}
A detailed analytical study of \eqref{eq:variance_r0} goes beyond the scope of this work.
Here, we limit ourselves to prove i) that in thermodynamic equilibrium $\text{Var}^\text{eq}\,\vec{r}_0 = k_BT/\kappa$ (corresponding to a null fluctuation enhancement, $\Delta_x=0$), and ii) that for small drag speed $\Delta_x=O(v^2)$.
Informally, the first point follows from the translational invariance of the canonical equilibrium distribution (conditioned on the position of the probe), leading to energy equipartition for the probe, while the second point follows from symmetry considerations, since the variance is left unchanged by switching to a reference frame where $\vec{v}'=-\vec{v}$. 

More formally, for the first point, note that thermodynamic equilibrium requires $v=0$.
The correlation $C=\sum_i\avg{\vec{r}_0\cdot \vec{F}_0(\vec{r}_{0i})} $ in~\eqref{eq:variance_r0} reads
\begin{align}
     C^\text{eq}=-N\int \d \vec{r}_0 P_p^{\text{eq}}(\vec{r}_0) \vec{r}_0\cdot \int \d \vec{r} \nabla {U}_0(\vec{r}_0-\vec{r}) \rho_f^{\text{eq}}(\vec{r}|\vec{r}_0)\,,
     \label{eq:eqm_correlation}
\end{align}
where we introduced the conditional equilibrium density of the units given the probe, which in equilibrium must be translationally invariant, \emph{i.e.},  $ \rho_f^{\text{eq}}(\vec{r}|\vec{r}_0) = \rho_f^{\text{eq}}(\vec{r}-\vec{r}_0)$.
Furthermore, $\rho_f^{\text{eq}}$ is also spherically symmetric (to guarantee reciprocity), and since $\vec{F}_0=-\nabla U_0$ where $U_0$ is a spherically symmetric interaction potential, the inner integral in \eqref{eq:eqm_correlation} vanishes.
Analogous considerations lead to $\avg{\vec{F}_0(\vec{r}_{0i})}_{\text{eq}}=0$.
Inserting both results in \eqref{eq:variance_r0} proves that $\Delta_x=0$ is in equilibrium.

To show that $\Delta_x=O(v^2)$ at small enough $v$, we consider the highly dilute limit, where the stationary fluid density $\rho_f(\vec{r}|\vec{r}_0)$ is well described by~\cite{van2024soft}
\begin{align}
    0 = \nabla \cdot \left[\left( - \varepsilon \vec{u}_\parallel - \frac{\epsilon_0}{k_BT}\nabla U_0 \right) \rho_f - \nabla\rho_f\right]\,.
\label{eq:fluid_density_dilute_varepsilon}
\end{align}
Eq.~\eqref{eq:fluid_density_dilute_varepsilon} has been written in adimensional form by introducing the rescaled variables (that we use throughout the remaining of this Section) $\vec{r}=\sigma_0 \vec{r}'$, $\vec{v}=v \vec{u}_\parallel$, $U_0 = \epsilon_0 U_0'$, $\rho_f'=\rho_f \sigma_0^2$, dropping the primes, and defining $\varepsilon = \gamma_0 v \sigma_0 / k_BT$.
We now solve this equation perturbatively up to the second order by introducing the expansion
\begin{align}
    \rho_f = \rho_f^{(0)} + \varepsilon \rho_f^{(1)}+O(\varepsilon^2)\,. \label{eq:pert_theory_rho_fluid}
\end{align}
At order $O(\varepsilon^0)$, the solution of Eq.~\eqref{eq:fluid_density_dilute_varepsilon} is $\rho_f^{(0)}(\vec{r}|\vec{r}_0) = \rho_{f}^{\text{eq}}(\vec{r}-\vec{r}_0)=Z_{f}^{-1}\exp\left\{- \epsilon_0 U_0(\vec{r}-\vec{r}_0)/k_BT\right\}$, which coincides with the equilibrium density.

To obtain the next order, we notice that the only scalar quantities compatible with invariance under inversion of the reference frame are the powers of $\vec{u}_\parallel \cdot (\vec{r}-\vec{r}_0)$ and $\vec{u}_\parallel \cdot \nabla U_0(\vec{r}-\vec{r}_0)$.
Employing the ansatz $\rho_f^{(1)} =  \rho_f^{(0)} \vec{u}_\parallel \cdot \left(A (\vec{r}-\vec{r}_0) + B  \nabla U_0(\vec{r}-\vec{r}_0)\right) $, we arrive at
\begin{align}
    \rho_f^{(1)} = - \rho_f^{(0)}(\vec{r}|\vec{r}_0)\vec{u}_\parallel \cdot (\vec{r}-\vec{r}_0)\,,
    \label{eq:pert_theory_rho_fluid_1}
\end{align}
which is only meaningful for $\abs{\vec{r}-\vec{r}_0}\ll  \varepsilon^{-1}$.

Now, we estimate the moments appearing in \eqref{eq:variance_r0}.
First, we notice that $\avg{\vec{r}_0}=O(\varepsilon)$ and $\avg{\vec{F}_0(\vec{r}_{0i})}=O(\varepsilon)$, since the zeroth order contribution vanishes at equilibrium by symmetry, as explained above.
Therefore, the only possible  contribution of order $O(\varepsilon)$ to the variance \eqref{eq:variance_r0} comes from the correlation $\sum_i\avg{\vec{r}_0\cdot \vec{F}_0(\vec{r}_{0i})} $.
The latter can be written as 
\begin{align}
\begin{split}
    C&=-N\int \d \vec{r}\d\vec{r}_0 \vec{r}_0 \cdot \nabla_{\vec{r}_0} U_0  \\
    &\quad \times \left( P_p^{(0)}(\vec{r}_0)\rho_f^{(1)}(\vec{r}|\vec{r}_0) + P_p^{(1)}(\vec{r_0})\rho_f^{(0)}(\vec{r}|\vec{r}_0) \right)\,,
\end{split}
\end{align} 
in which the probability $P_p^{\varepsilon}(\vec{r}_0)$ of the probe appears.
Its equilibrium limit is the zero-mean Gaussian $P_p^{(0)}(\vec{r}_0)=P_p^{\text{eq}}(\vec{r}_0)=Z_p^{-1}\exp\left\{-\kappa\sigma_0^2\vec{r}_0^2/2k_BT\right\}$.
The contribution containing $P_p^{(1)}$ vanishes by symmetry as in Eq.~\eqref{eq:eqm_correlation}.
The remaining one can be evaluated using Eqs.~\eqref{eq:pert_theory_rho_fluid} and \eqref{eq:pert_theory_rho_fluid_1}, yielding (after the change of variables $\vec{r}'=\vec{r}-\vec{r}_0$)
\begin{align}
\begin{split}
    C&=N\int \d \vec{r}' \rho_f^{\text{eq}}(\vec{r}')  \left(\vec{u}_\parallel \cdot \vec{r}'\right)   \\
    &\quad\times \nabla_{\vec{r}'} U_0(\vec{r}') \cdot\int \d\vec{r}_0 P_p^{(0)}(\vec{r}_0) \vec{r}_0 =0\,,
    \end{split}
\end{align}
which that proves (for a generic, centrally-symmetric potential $U_0$) $ \mathrm{Var}\, \vec{r}_0 = k_BT/\kappa \sigma_0^2 + O(\varepsilon^2)$, or equivalently $\Delta_x =O(v^2)$.

\section{Far-from-equilibrium dynamics}
\label{sec:hopping}

\begin{figure*}
    \centering
    (a)
    \includegraphics[width=\textwidth]{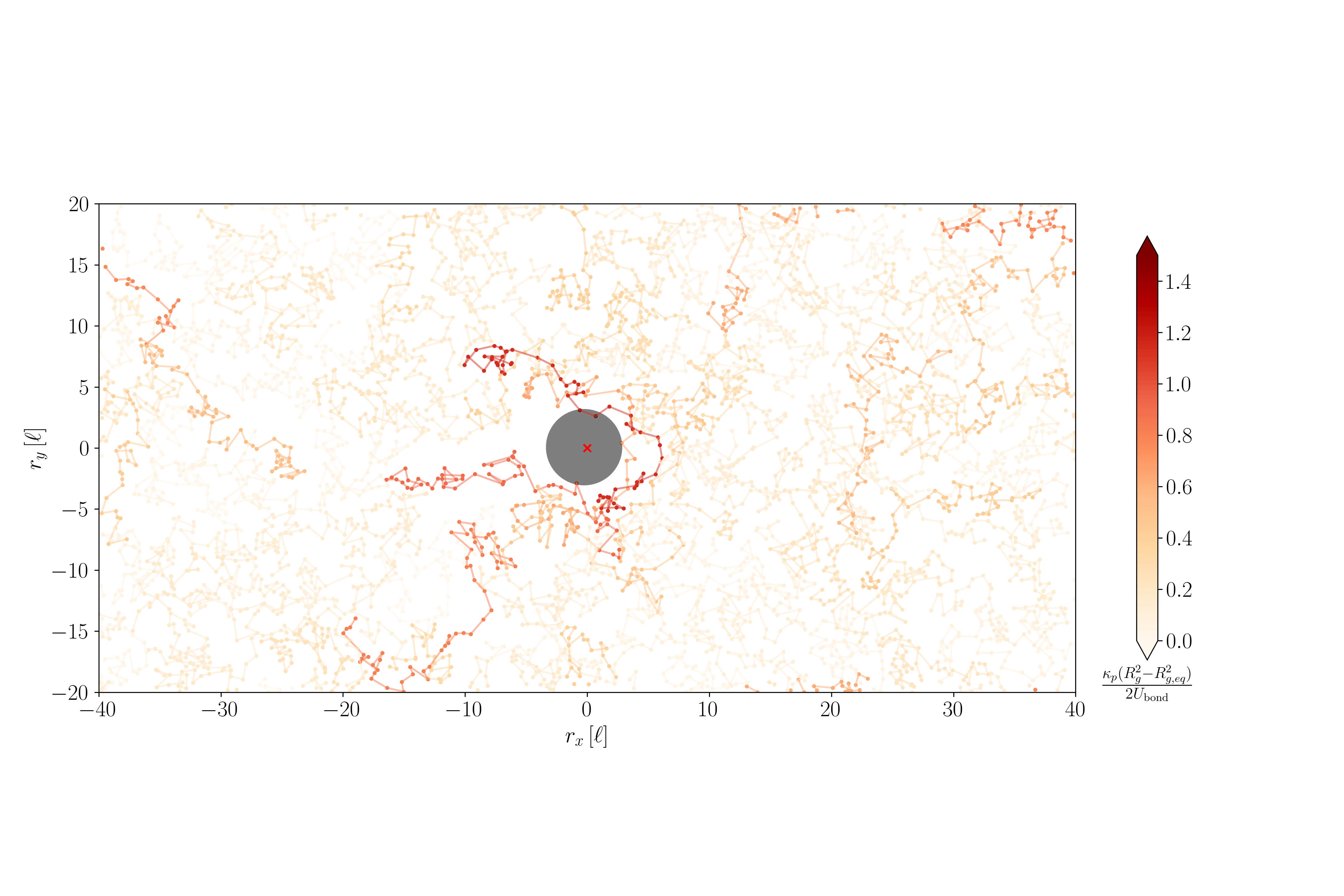}
    \vskip 2mm

    (b)\\
    \begin{tikzpicture}
\begin{axis}[
    width=8cm,
    height=4.5cm,
    xmin=0.5,
    xmax=8,
    ymin=-1,
    ymax=2,
    axis equal image,
    axis y line=none,
    axis x line=bottom,
    xtick=\empty,
    ytick=\empty,
]
\addplot[domain=-180:180,samples=100,thick] ({6+cos(x)},{sin(x)});
\addplot[domain=1.5:3.5,samples=100,
] {1 + 0.15*sin(deg(10*x))};
\addplot[domain=3.5:6,samples=100,
] {1 + 0.1*sin(deg(10*x))};
\fill[black] (axis cs:1.5,1) circle (6pt) node[above=5pt]{$r_3$};
\fill[black] (axis cs:3.5,1) circle (6pt) node[above=5pt] {$r_2$};
\fill[black] (axis cs:6,1) circle (6pt) node[above=5pt]{$r_1$};
\draw[->, line width=1.5pt] (axis cs:1.5,1) -- (axis cs:0.6,1) node[midway,above] {$\gamma v$};
\draw[->, line width=1.5pt] (axis cs:3.5,1) -- (axis cs:2.6,1) node[midway,above] {$\gamma v$};
\end{axis}
\end{tikzpicture}
    \caption{(a) Typical configuration at high speed, obtained for $L=56$ and $\mathcal{U}=0.455$. 
    The color bar shows the excess elastic energy of the chains (compared to the equilibrium value) in units of $U_\text{bond}$. Note various chains hanging around the probe moving to the right (b) A scheme of the one-dimensional model used to compute the elastic energy of half of the chain right before detachment.} 
    \label{fig:hooked_polymer}
\end{figure*}

\begin{figure*}
\centering
\includegraphics[width=0.5\linewidth]{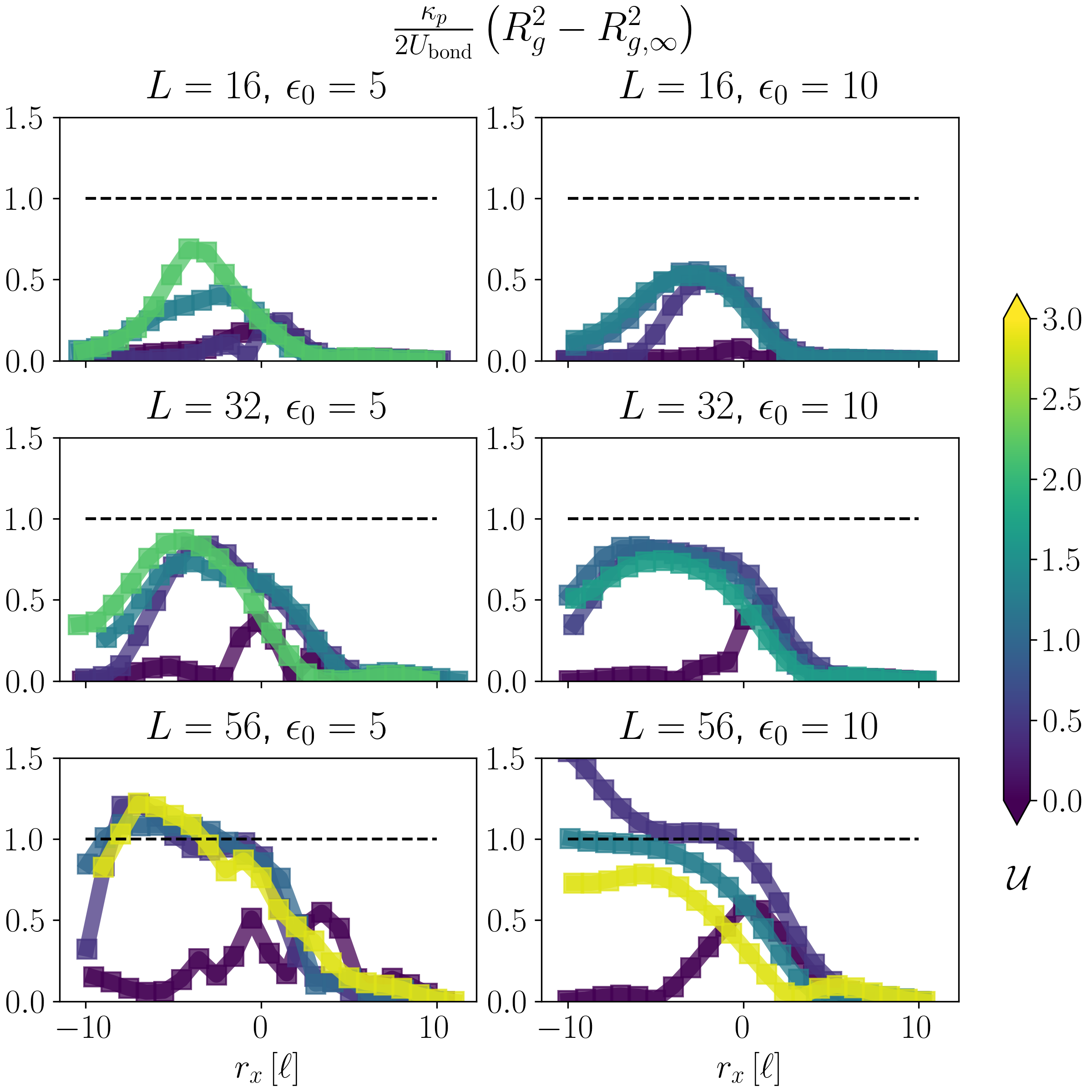}
    \caption{Profile of the excess elastic energy with respect to the equilibrium one, in chains around the probe (along the $x$ axis of motion of the trap). The three panels on the left are for different $L$'s and for $\epsilon_0=5 \mathcal{E}$, which is half of the typical $\epsilon_0=10 \mathcal{E}$ we adopted for the simulations in this work (right panels).}
\label{fig:excess_elastic_energy_epsilon0}
\end{figure*}

We focus on the situation in which, at sufficiently high velocity, a single chain is trapped in the typical configuration shown in Fig.~\ref{fig:hooked_polymer}(a) by the probe, for which we develop the simplified model in Subsection \ref{sec:jump_process}.
In Subsection \ref{sec:elastic_energy}, we estimate the critical velocity at which activated jump events become relevant by considering a simplified one-dimensional setting, depicted in Fig.~\ref{fig:hooked_polymer}(b).
Furthermore, we show that this activated process results in the relaxation of the elastic stress accumulated by the probe when $\mathcal{U}\geq 1$ by looking at the amount of excess elastic stress that can be accumulated near the probe (compared to the equilibrium one), as shown in Fig.~\ref{fig:excess_elastic_energy_epsilon0} for various choices of length $L$, rescaled velocity $\mathcal{U}$ and repulsive energy scale $\epsilon_0$ of the probe.

\subsection{Hopping process at large velocities.}
\label{sec:jump_process}

In our model, a bond of a chain, perturbed by the traveling probe, may stretch so much that the probe can pass through it.
In our two-dimensional model, this kind of event exemplifies drastic, three-dimensional rearrangements of the polymers. 
Such events should occur when the polymers' diffusive or reptation dynamics become too slow to escape from the region in front of the fast-approaching probe, so that entangled polymers progressively accumulate stress, which is released only when a critical threshold is reached.

The simplest configuration that allows for an explicit estimate of the threshold energy is as follows:
Two units at the ends of the stretched bond occupy two positions $\pm y$ at $x=0$, yielding a total energy as a function of $y$
\begin{align}
    U(y) &= 2 U_0(y) + U_b(2y)\nonumber\\
    &=2\left[
    \epsilon_0 e^{-y^2/2\sigma_0^2} + \kappa_p y^2
    \right]
    \label{eq:U}
\end{align}
where $U_b(2y)$ is the elastic energy of the bond.
For simplicity, we are neglecting the repulsion between the units.
As Fig.~\ref{fig:U} shows, $U(y)$ has a single minimum at $y=y^*\equiv \sqrt{2\ln (\epsilon_0/(2\kappa_p\sigma_0^2))} \sigma_0$,
\begin{align}\label{eq:Ubond}
U_\textrm{bond} \equiv U(y^*) = 2\kappa_p \sigma_0^2 \left[1 - 
   \ln \left(\frac{2 \kappa_p \sigma_0^2}{\epsilon_0} \right) \right].
\end{align}
With the parameters of our simulations, the threshold energy is $U_\textrm{bond} \approx 30 k_B T$. Hence, the stretching of a bond in the vicinity of the probe would be highly unlikely in thermal equilibrium. 
However, it becomes facilitated by the nonequilibrium force arising from the relative motion of the units against the probe at sufficiently large $v$, as will be discussed in Subsection \ref{sec:elastic_energy}.

\begin{figure*}
\centering
\begin{tikzpicture}
\begin{axis}[
    width=12cm,
    height=4cm,
    xlabel={$y\, [\ell]$},
    ylabel={$U/k_BT$},
    xmin=0,
    xmax=6,
    ymin=0,
    ymax=50,
    grid=major,
    legend pos=north east,
    declare function={
        e0 = 20;
        sigma0 = 11/8;
        kp = 2;
        xmin = sigma0*sqrt(2*ln(e0/(2*kp*sigma0^2)));
        f1(\x) = 2*e0*exp(-\x*\x/(2*sigma0^2));
        f2(\x) = 2*kp*\x*\x;
        ftot(\x) = f1(\x) + f2(\x);
    },
]
\addplot[blue, dashed, thick, domain=0:4, samples=200] {f1(x)};

\addplot[red, dotted, thick, domain=0:4] {f2(x)};

\addplot[black, solid, thick, domain=0:4, samples=200] {ftot(x)};

\fill[black] (axis cs:{xmin},{ftot(xmin)}) circle (2pt) node[above=5pt]{$U_{\textrm{bond}}$};

\legend{
    {$2 U_0(y)$},
    {$U_b(2y)$},
    {$2 U_0(y) + U_b(2y)$}
}

\end{axis}
\end{tikzpicture}
\caption{Potential energy \eqref{eq:U} of two joined units stretched at a distance $2y$ on the two sides of the probe (black line) and its two contributions from the probe repulsion ($2 U_0(y)$) and the bond elastic energy ($U_b(2y)$).}
\label{fig:U}
\end{figure*}

Under such conditions, we now introduce a minimalistic model describing the opening and sudden closing of the bond between two units in the chain. 
This activated process displaces a portion of the chain from the front of the probe to its wake.
Such portion involves a few units, and while it might depend on the probe's effective radius $ y^* \approx 2\sigma_0$, it should not depend sensibly on the chain length $L$.

For dynamics \eqref{eq:langevin}, we focus on large velocities $\vec{v}$ and long chains ($L\gg 1$) and assume that a single chain hooked and stretched by the probe is the configuration that contributes the most to the probe's fluctuations. 
From \eqref{eq:monomer_comoving_langevin}, noting that the interactions between units of the same chains sum up to zero and neglecting the interactions with other chains, we get the dynamics of the chain's center of mass ${\vec{r}}_\mathrm{cm}= \sum_{i=1}^L \vec{r}_i /L$,
\begin{align}
   L \gamma \dot{\vec{r}}_\mathrm{cm} &=  -L \gamma \vec{v} -  \vec{F}_{\mathrm{P} 0 }
    +\sqrt{2 L\gamma k_BT}   \vec{\xi}_{\rm cm} ,\label{eq:cm_polymer}
\end{align}
where we defined the force of the units on the probe as $\vec{F}_{\mathrm{P} 0 } \equiv \sum_{i=1}^L \vec{F}_0(\vec{r}_{0i})$.
By plugging it into the equation for the probe, valid under the same approximation, we get
\begin{align}
    \gamma_0\dot{\vec{r}}_0& =  -\gamma_0\vec{v}  -\kappa\vec{r}_0+\vec{F}_{\mathrm{P} 0 } +\sqrt{2\gamma_0k_BT}\vec{\xi}_0\, . 
\end{align}
Defining further $\Gamma=(\gamma_0 + L \gamma)$, for the component $x_0=\vec{r}_{0,x}\cdot\vec{u}_\parallel$ along the drag direction, we obtain
\begin{align}
    \gamma_0\dot x_0 & =  -\Gamma v  -\kappa x_0 - L \gamma \dot x_\mathrm{cm} +\sqrt{2\Gamma k_BT} \xi\,.
    \label{eq:r0simpl}
\end{align} 
We assume that negative increments of the center of mass ${x}_\mathrm{cm}$ occur with jumps driven by a Poisson process $\eta$ with hopping rate $\varphi(v)$. 
As argued above, such jumps involve a finite, $L$-independent fraction of units that relocate from the front to the back of the probe.
Hence, each of these events generates a positive impulse $\mathcal{I}$ on $x_0$ that is expected to be independent of $L$ and $v$, but possibly a function of the energy scale $U_\mathrm{bond}$. 
We thus write $- L \gamma \dot x_\mathrm{cm}=\mathcal{I} \eta$, which renders \eqref{eq:r0simpl} in the form
\begin{align}
    \gamma_0\dot x_0 & =  -\Gamma v  -\kappa x_0 +\mathcal{I}\, \eta +\sqrt{2\Gamma k_BT}\xi \,,
\end{align}
whose associated master equation is
\begin{align}
\begin{split}
    \partial_t p(x_0,t) =&
-\partial_{x_0} \left[ -\gamma_0^{-1} (\Gamma v + \kappa x_0) p(x_0,t) \right.\\
&\left. \quad  -\gamma_0^{-2}  \Gamma k_BT \partial_{x_0}  p(x_0,t) \right] \\
&\quad + \varphi \left[p(x_0-\mathcal{I}/\gamma_0 ,t) - p(x_0,t)\right].
\end{split}
 \end{align}
It can be written in terms of the characteristic function $g(q)\equiv \avg{e^{i q x_0}}$ in the steady state as
\begin{align}
\begin{split}
    0 &= 
i q [ -\gamma_0^{-1}  (\Gamma v -i  \kappa \partial_q) +i q \gamma_0^{-2}  \Gamma k_BT ]g(q)\\
&\quad + \varphi (e^{iq \mathcal{I}/\gamma_0}   - 1)  g(q),
\end{split}
 \end{align} 
which gives the cumulant generating function $K(q)\equiv \ln g(q)$ (with $\mathrm{Ein}$ the complementary exponential integral~\cite{NIST:DLMF}),
\begin{align}
 K(q)=\frac{\gamma_0}{\kappa}
 \left[
 -\frac{\i  \mathcal{I} \varphi}{\gamma_0}
 \,\mathrm{Ein}
 \left( -\frac{\i q \mathcal{I}}{\gamma_0}\right)
 -\frac{\Gamma k_B T q^2}{2\gamma_0^2}  -\i \frac{\Gamma}{\gamma_0} q v 
 \right].
\end{align}
The mean and the variance are, respectively,
\begin{align}
\avg{x_0} &= -i \lim_{q \to 0} \partial_{q} K= -\frac{\Gamma  v}{  \kappa \gamma_0 } + \frac{\mathcal{I}}{ \kappa} \varphi  \,,\\
\text{Var}\, x_0&=-\lim_{q \to 0} \partial_{q}^2 K = \frac{\Gamma  k_B T \mu }{\kappa }+\frac{\mathcal{I}^2 \varphi}{2 \gamma_0 \kappa}\,.
\end{align} 
The ratio between the variance and the mean, compared to the case $\varphi(v=0)=0$ in equilibrium, is
   \begin{align}
&\frac{\text{Var}\, x_0-\text{Var}_\mathrm{eq} x_0}{|\avg{x_0}|}=\frac{\varphi\mathcal{I}^2}{2 \gamma_0 |\varphi \mathcal{I}-\Gamma  v|}.
  \end{align} 
We see that the two quantities are proportional to each other when the hopping process is dominant, and their ratio is independent of $L$ and $v$,
\begin{align}
&\frac{\text{Var}\, x_0-\text{Var}_\mathrm{eq} x_0}{|\avg{x_0}|}\underset{\varphi \mathcal{I} \gg \Gamma v}{\longrightarrow} \frac{\mathcal{I}}{2 \gamma_0}.
\end{align}
This plateau is clearly visible in Fig.~3(c) of the main text, and the activated jumps provide a mechanism for the elastic stress relaxation visible in Fig.~\ref{fig:excess_elastic_energy_epsilon0} for $L=32, 56$ and $\mathcal{U}>1$, in the region $r_x>0$ (corresponding to the front of the probe).

\subsection{Threshold velocity from relaxation of elastic stress}
\label{sec:elastic_energy}

We derive a theoretical estimate $v^*_\text{th}$ of the threshold velocity $v^*$ that marks the crossover from the advected-dominated regime to the hopping-dominated one. We assume that the hopping rate $\varphi$ follows the Arrhenius law
\begin{align}
    \varphi(v) \approx e^{-\beta(U_\text{bond}-U_\text{el}(v))},
\end{align}
where $U_\text{el}(v)$ is the elastic energy accumulated in the hooked chain due to the probe motion at mean speed $v$.
The threshold velocity corresponds to the value $v^*_\text{th}$ at which $\varphi \approx 1$, namely, 
\begin{align}\label{eq:theor_v^*}
   U_\text{el}(v^*_\text{th}) = U_\text{bond}.
\end{align}

The elastic energy $U_\text{el}(v)$ is calculated as follows.
If the system is sufficiently dilute and the stretching of the chain resulting from the motion of the probe is significant, in a first approximation we can neglect the repulsive interaction between units.
As a result, we can describe the motion of the units in the single hooked chain in the frame where the probe is at rest according to the following Langevin equation
\begin{align}
\begin{split}
        \dot{r}_i &= -v + \gamma^{-1}{F}_0(\vec{r}_{i0}) \\
    &\quad +  \gamma^{-1}\kappa_p  \sum_{j=2}^{L'-1} (\delta_{j,i+1} + \delta_{j,i-1})({r}_{j}-r_i)+  \sqrt{2D}{\xi}_i\,,
\end{split}
\end{align}
for $i=1, \dots, L'$, where $L'<L$ is the number of units in one of the two lateral sub-chains of a hooked chain, such as the darkest one in Fig.~\ref{fig:hooked_polymer}(a). 
One sub-chain is also sketched in Fig.~\ref{fig:hooked_polymer}(b).
Furthermore, we assume that the force exerted by the probe is such that the velocity of the first units is fixed to zero, \emph{i.e.} $\dot{r}_1 = 0$, while it is negligible for the units with index $i\geq 2$.

Supposing that the chain remains hooked for a duration sufficient to allow the entire chain to reach a steady state, the problem can be mapped onto an equilibrium scenario. In this case, the first unit is pinned at $r_1=0$. In contrast, each remaining unit of the elastic chain is subject to a uniform force $\gamma v$, resulting from the potential energy $U_{\gamma}(r_i) = - \gamma v r_i$.
The total potential energy is the sum of the linear potential $U_{\gamma}$, resulting from the motion of the probe, plus the elastic potential energy of the chain; it can be written in terms of the adimensional bond length $\Delta r_i \equiv (r_{i+1}-r_i)/\ell$, (implying $r_i = \ell \sum_{k=1}^{i-1}\Delta r_k$), 
\begin{align}
\begin{split}
        U_{\text{tot}} 
    &= \frac{1}{2}\kappa_p\ell^2 \sum_{i=1}^{L'-1} \Delta r_i^2 - \gamma v \sum_{i=2}^{L'} r_i\\
    &=  \sum_{i=1}^{L'-1}  \underbrace{\frac{1}{2}\kappa_p \ell^2 \Delta r_i^2 - \gamma v \ell (L'-i) \Delta r_i}_{\equiv U_i}\,.
\end{split}
\end{align}
The corresponding partition function factorizes in terms of single-bond partition functions $Z = \prod_{i=1}^{L'-1} Z_i$. 
The partition function $Z_i$ reads
\begin{align}
\begin{split}
 Z_i &= \int \d \Delta r_{i} \exp\{-\beta U_{i}\} \\
            &= 
    \exp\left\{\frac{\beta  \gamma^2 v^2 (L'-i)^2}{2\kappa_p } \right\} \\
    &\quad \times
    \int \d \Delta r_{i}  \exp \left\{- \frac{1}{2}\beta \kappa_p \ell^2 \left(\Delta r_i - \frac{\gamma v (L'-i)}{\kappa_p\ell}\right)^2 \right\}\\
    &= \sqrt{\frac{2\pi}{\beta\kappa_p \ell^2}}\exp\left\{\frac{\beta  \gamma^2 v^2 \left( L'^2 - 2 L'i + i^2\right) }{2\kappa_p } \right\} \,,
\end{split}
\end{align}
where the last equality follows from Gaussian integration.
The free energy of the system is 
\begin{align}
\begin{split}
        F &= -\beta^{-1} \ln Z \\
    &= -\beta^{-1} \sum_{i=1}^{L'-1} 
    \left[ -\frac{1}{2}\ln\frac{\kappa_p\ell^2\beta}{2\pi} 
    + \frac{\beta \gamma^2 v^2 (L'^2-2L'i+i^2)}{2\kappa_p}\right]\\
    &= 
    \frac{L'-1}{2\beta}\ln\frac{\kappa_p\ell^2\beta}{2\pi} 
    +\frac{\gamma^2 v^2 L'^3}{2\kappa_p}
    \left(\frac{3}{2}-2L'^{-1} +\frac{1}{2}L'^{-2}
    \right)\,.
\end{split}
\end{align}
In the last equality, we used the finite sums $\sum_{k=1}^{N} k= N(N+1)/2$ and $\sum_{k=1}^{N} k^2= N(N+1)(N+1/2)/6$, which can be obtained from the generating function $g_N(q)=\sum_{k=1}^N \e^{- q k}= (1-\e^{-qN})/({\e^{q}-1})$.
The average elastic energy stored in the chain  is 
\begin{align}\label{eq:Uel}
    U_{\textrm{el}} = 2(- \partial_\beta \ln Z) = \frac{\gamma^2 v^2 L'^3}{\kappa_p}
    \left(\frac{3}{2}-2L'^{-1} +\frac{1}{2}L'^{-2}
    \right)\,.
\end{align}
The factor $2$ comes from assuming that each of the sub-chains of a hooked chain in the configuration in Fig.~\ref{fig:hooked_polymer}(a) can be represented by the simplified setup of Fig.~\ref{fig:hooked_polymer}(b).

Therefore, plugging \eqref{eq:Uel} into \eqref{eq:theor_v^*} and approximating $L'$ by $=L/2$ we arrive at 
\begin{align}
    v^*_\text{th} = \frac{ \sqrt{2 \kappa_p U_\text{bond}} }{\gamma} \left(\frac{3}{8} L^{3} - L^2+ \frac{1}{2}L  \right)^{-1/2},
    \label{eq:vstartheory}
\end{align}
which explains the empirically determined scaling $v^*_\text{th} \sim v^* \sim L^{-3/2}$, for large $L$.

\bibliography{rsc}

\end{document}